\documentclass[12pt]{article}

\usepackage{amssymb, amsmath}
\usepackage{setspace} 
\begin{document} 

 
\title{Singularity confinement for 
maps with the Laurent property}

\author{ 
A.N.W. Hone\thanks{  
Institute of Mathematics, Statistics \& 
Actuarial Science,  
University of Kent,  
Canterbury CT2~7NF, United Kingdom  }  
} 

 
\def\underset#1#2{\mathrel{\mathop{#2}\limits_{#1}}}  
\newcommand{\haf}{{\hat{f}}}  
\newcommand{\beq}{\begin{equation}} 
\newcommand{\eeq}{\end{equation}} 
\newcommand{\bea}{\begin{eqnarray}} 
\newcommand{\eea}{\end{eqnarray}} 
\newcommand\la{{\lambda}} 
\newcommand\ka{{\kappa}} 
\newcommand\al{{\alpha}} 
\newcommand\be{{\beta}} 
\newcommand\de{{\delta}}
\newcommand\si{{\sigma}} 
\newcommand\lax{{\bf L}} 
\newcommand\mma{{\bf M}}  
\newcommand\ctop{{\mathcal{T}}}  
\newcommand\hop{{\mathcal{H}}}  
\newcommand\ep{{\epsilon}}
\newcommand\T{{\tau}}
\newcommand\om{{\omega}}
\newcommand\ga{{\gamma}}
 
\newcommand{\N}{{\mathbb N}}  
\newcommand{\Q}{{\mathbb Q}}  
\newcommand{\Z}{{\mathbb Z}}  
\newcommand{\C}{{\mathbb C}}
\newcommand{\R}{{\mathbb R}}
 
\newtheorem{Pro}{Proposition} 
\newtheorem{Lem}{Lemma} 
\newtheorem{The}{Theorem} 
 
\newcommand{\rd}{{\mathrm{d}}} 
\newcommand{\sgn}{{\mathrm{sgn}}} 
\def \ring {{\cal R}} 

\maketitle 
 
\begin{abstract}  
The singularity confinement test 
is very useful  
for isolating integrable cases of discrete-time 
dynamical systems,   
but it does not provide a 
sufficient criterion for integrability. 
Quite recently a new property of 
the bilinear equations 
appearing in discrete soliton theory has been noticed: the iterates 
of such equations are Laurent polynomials in the initial data. 
A large class of non-integrable mappings of the plane are presented 
which both 
possess    
this Laurent property and have confined
singularities. \textbf{MSC2000 numbers: 11B37, 93C10, 93C55}  
\end{abstract}  


 
There continues to be a great deal of interest in discrete-time 
dynamical systems that are integrable. There is a vast range of 
such systems, including symplectic maps 
and B\"acklund transformations for Hamiltonian systems in classical mechanics 
\cite{rag}, mappings that preserve plane curves \cite{qrt}   
which occur in statistical mechanics, discrete 
analogues of Painlev\'e transcendents \cite{dpainleve}, partial 
difference soliton equations appearing in numerical 
analysis and solvable quantum  models \cite{quant}, and equations 
arising in theories of discrete geometry and discrete analytic functions 
\cite{digp}.  For some time it has been appreciated that 
integrability  in the discrete setting is associated with 
certain weak growth phenomena \cite{ves}, which can be measured 
by means of suitable notions of entropy \cite{hv} or complexity \cite{aabhm}. 

Given the multitude of 
application areas in which discrete integrable systems 
appear, the problem of identifying when a given system is 
integrable is of considerable importance. In the continuous setting, 
the Painlev\'e property has proved to be an extremely useful criterion for 
isolating  integrable differential equations \cite{conte}, and this led 
Grammaticos, Ramani and Papageorgiou to introduce the 
singularity confinement test
for discrete equations \cite{grp}. However, while that test 
has been enormously successful at identifying discrete 
Painlev\'e equations, it turns out that singularity 
confinement is not sufficient for integrability, as was 
pointed out by Hietarinta and Viallet \cite{hv}. Those authors 
found numerous examples of rational maps, taking the form  
\beq \label{addmap} 
x_{n+1}+ x_{n-1}=f(x_n),  
\eeq 
which have confined singularities and yet whose orbit 
structure displays the characteristics of chaos. This led 
them to suggest that the stronger requirement of 
zero algebraic entropy (defined 
in terms of the growth of degrees of iterates) should be 
a necessary property of rational maps that are integrable, 
in agreement with the observations of Veselov \cite{ves}.  
More recently it was proposed by Ablowitz, Halburd and Herbst 
that the Painlev\'e property can be extended to difference equations
using Nevanlinna theory \cite{ahh}, 
while Roberts and Vivaldi have used the orbit structure of 
rational
maps defined over finite fields to detect integrability 
\cite{rv}, and  
Halburd has 
translated the concepts of \cite{ahh}
into a Diophantine integrability
criterion for discrete equations \cite{halburd}.
  
There is a large amount of literature on discrete bilinear equations, 
including the
bilinear forms of discrete
Painlev\'e  equations \cite{dpainleve}, and
bilinear partial difference equations such as
Hirota's difference equation \cite{quant}. 
However, there is one aspect of such integrable bilinear equations that 
researchers on integrable systems have apparently overlooked, 
namely the fact that they have the \textit{Laurent property}: that is, for 
suitably specified initial data,  
all of the iterates of these discrete equations 
are Laurent polynomials in these  data
with integer 
coefficients. 
It seems that this Laurent phenomenon 
was originally known only to a few people working in 
algebraic combinatorics,  
and for Hirota's equation 
it was first proved by Fomin and Zelevinsky within the
framework of their theory of cluster algebras \cite{fz},
with further 
combinatorial interpretations being found later  
\cite{propp1}. 

One of the simplest examples of a bilinear equation displaying the 
Laurent  phenomenon
was found by Michael Somos, who considered 
$k$th order recurrences of the form
\beq  \label{somos}
\tau_{n+k}\tau_n=\sum_{j=1}^{[k/2]}\tau_{n+k-j}\tau_{n+j}, \qquad k\geq 4,  
\eeq
taking the initial values $\tau_0=\tau_1=\ldots =\tau_{k-1}=1$.
Clearly each new iterate $\tau_{n+k}$  
of (\ref{somos})  
is a rational function of the initial data, so
one expects $\tau_n\in\Q$, 
but it was observed numerically that for the Somos-4 recurrence
\beq
\tau_{n+4}
\tau_{n}
=\al \,
\tau_{n+3}
\tau_{n+1}
+\beta
\,
(\tau_{n+2})^2
\label{bil}
\eeq
with parameters $\alpha=\beta=1$ and starting with four ones, an
integer sequence
$$1,1,1,1,2,3,7,23,59,314,1529, 8209, 83313,\ldots$$  
results. 
Several simple proofs of the integrality  
of this sequence were subsequently obtained  
(see the article by Gale and the other references \cite{wardetal}). 
However, it was   
realized that the deeper reason behind this lay in the fact that the
recurrence (\ref{bil}) has the Laurent property, meaning that 
the iterates are
polynomials in the coefficients, the four initial data, and their
inverses, and these Laurent polynomials have integer coefficients,
i.e. $\tau_n\in \mathbb{Z}[\al , \be , \tau_0^{\pm 1},
\tau_1^{\pm 1},  \tau_2^{\pm 1},  \tau_3^{\pm 1}]$ for all $n$. 

Fomin and Zelevinsky found that their theory of cluster algebras 
provided a suitable setting within which
they could prove
the Laurent property for a variety
of discrete equations \cite{fz}, including  certain recurrences of the form
$$\tau_{n+k}\tau_n=F(\tau_{n+1},\ldots,\tau_{n+k-1})$$  
for particular choices of polynomials $F$ (including 
Somos-$k$ for $k=4,5,6,7$),   
as well as integrable two- and
three-dimensional bilinear recurrences like Hirota's equation. 
As a parallel development, 
the connection between the iterates of the 
general Somos-4
recurrence (\ref{bil})  and sequences of points on
elliptic curves has been explained by several people: 
two different approaches are found respectively 
in the PhD thesis of Swart 
and in the work of van der Poorten  
\cite{vdp}, while independently the author found
the explicit solution of the initial value problem
for both Somos-4 and Somos-5 \cite{honetams} in terms
of elliptic sigma functions. For earlier unpublished results of 
Zagier and Elkies on the associated 
elliptic curve and theta function formulae for  
the original Somos-5 sequence, see 
\cite{zagier} and \cite{heron} respectively.

An essential observation in the work  
\cite{honetams} 
was that both the fourth and the fifth order 
Somos recurrences 
could be understood in terms of a suitable
integrable mapping of the plane. For example, 
setting $x_n=\tau_{n+1}\tau_{n-1}/(\tau_n)^2$ transforms 
(\ref{bil}) into 
the rational map 
\beq 
\label{imap} 
x_{n+1}\, x_{n-1}= 
\frac{\al }{ x_n}  
+\frac{\be }{x_n^2}  
\eeq  
which preserves the two-form $(dx_{n-1}\wedge dx_n)/(x_{n-1}x_n)$ and 
has the conserved quantity 
\beq 
\label{curve} 
J=x_{n-1}x_n +\al\left( 
\frac{1}{x_{n-1}} +\frac{1}{x_{n}} 
\right) + 
\frac{\be}{x_{n-1}x_n}.  
\eeq
A symplectic map of the plane with  a conserved 
quantity is the discrete analogue of a Hamiltonian 
system with one degree of freedom, and hence is  
integrable in the sense of Liouville (see chapter 10 in \cite{arnold}). 
Thus the map defined by (\ref{imap}) is integrable and belongs 
to the well 
studied class \cite{qrt, ves} of rational mappings of the plane 
that preserve an  
algebraic curve, in this case 
the curve of genus one defined by (\ref{curve}). 
In fact, certain integer sequences satisfying 
the Somos-4 recurrence (\ref{bil}) were already known 
to number theorists by the name of elliptic divisibility 
sequences \cite{wardetal}, and these continue to be the subject of 
active research due to the way that new prime divisors appear therein. 

The main result proved below is the following 

\noindent \textbf{Theorem} 
{\it Given a polynomial $f(x)$ of degree $d$ having the 
form 
$$ 
f(x)=x^M\, F(x) 
$$ 
for a non-constant polynomial $F$ with $F(0)\neq 0$, 
the recurrence  
\beq \label{prods} 
x_{n+1}\,x_{n-1}=f(x_n)   
\eeq 
possesses the Laurent property if and only if   
one of the following three cases holds:  

\noindent (i) $M=0$ and, for all $x$, the polynomial $f$ satisfies 
\beq
\label{recip}
f(x) =\pm x^d\, f(1/x), \quad  \mathrm{with}\,\, f(0)=1 \,\,
\mathrm{for}\,\, d\neq 2;
\eeq 

\noindent (ii) $M=1$ and, for all $x$, the polynomial 
$F$ (of degree $D$) satisfies 
\beq
F(x)=\pm x^D\, F(1/x), 
\quad  \mathrm{with}\,\, F(0)=1 \,\, \mathrm{or} \, -1 
\,\,\mathrm{for}\,\, D\neq 1; 
\label{recip2}  
\eeq

\noindent (iii) $M\geq 2$ and $F$ is an 
arbitrary non-constant polynomial. 
 
Moreover, up to the freedom to rescale $x_n$, in cases 
(i) and (ii)  
the respective 
conditions (\ref{recip}) and (\ref{recip2}) are also necessary and 
sufficient for the singularities of  
(\ref{prods}) to be 
confined immediately, while in case (iii) the singularity confinement 
test is failed. } 

Clearly to obtain Laurent 
polynomials it is necessary for 
$f(x)$ to be a polynomial in $x$, and the recurrence (\ref{prods}) 
is generated by iterating the rational mapping of the plane defined by  
\beq 
\label{phimap} 
\phi : \qquad \left( 
\begin{array}{c} x \\   y \end{array}  
\right)  
\mapsto  
\left( 
\begin{array}{c} y \\  f(y)/x  \end{array}  
\right), 
\eeq 
which preserves the two-form 
\beq 
\label{2form} 
\om = (xy)^{-1}\, dx\wedge dy.  
\eeq 
For (\ref{prods}) to have the Laurent property it is  required 
that $x_n\in \ring :=\Z [{\bf c}, x_0^{\pm 1},x_1^{\pm 1}]$ for all $n$, 
where $\bf{c}$ denotes the coefficients of $f$. To begin with, we  
take the generic case $f(0)=\lambda\neq 0$. Given the two initial data 
$x_0$, $x_1$, the next iterates are 
$x_2=f(x_1)/x_0$ and   
$x_3=f(x_2)/x_1$,  
which are both in $\ring$ for any $f$. The first place where 
a division must occur is for $x_4=f(x_3)/x_2$: if $x_4\in\ring$ then 
$f(x_1)=x_2\,x_0$ must divide $f(x_3)$. Now every Laurent 
polynomial  
can be written as a polynomial divided by a monomial $x_0^\ell x_1^m$, and  
the ring $\ring$ 
has the structure of a unique factorization domain, 
with units given by the monomials $\pm x_0^\ell x_1^m$ for 
integers $\ell,m$. Thus one can do 
modular arithmetic with the elements of $\ring$ in the usual way, 
and note that $x_1$ and $x_2$ are coprime. Then $x_3\equiv 
f(0)/x_1\equiv \la /x_1 \bmod x_2$ and 
hence $f(x_3)\equiv f(\la /x_1 ) \bmod x_2$, 
so for divisibility of $f(x_3)$ by $f(x_1)$ we must have that, for 
arbitrary  $x_1$,  
\beq 
\label{divcon} 
\la f(x_1)=\mu \, x_1^d \, f(\la /x_1 ) 
\eeq 
where $d=\deg f$ and the constant 
$\mu$ is the leading coefficient of $f$. 
Comparing the constant 
term on each side of (\ref{divcon}) gives $\la^2=\mu^2\la^d$, so 
$\mu =\pm \la^{1-d/2}$. If $d\neq 2$ then to have polynomials 
in $\la $ forces the choice $\la =1$ (and in any case, the freedom 
in $\la$ can be removed by rescaling $x_n$), so we find 
$\mu=\pm 1$ for all $d$, and thus the necessary condition for 
the recurrence (\ref{prods}) to have the Laurent property is that 
the polynomial $f$ satisfies the reciprocal property  (\ref{recip}). 

The condition (\ref{recip}) also turns out  to be sufficient 
to ensure that $x_n\in\ring$ for all $n$, and also implies 
that (\ref{prods}) passes the singularity confinement test. 
To see that this is sufficient for the Laurent property, 
one can apply Fomin and 
Zelevinsky's Caterpillar Lemma \cite{fz}, but 
a direct proof is given here 
for completeness. 
Take as the inductive hypothesis that 
$x_j\in\ring$ for $0\leq j\leq n$ with all adjacent pairs 
$x_j$, $x_{j+1}$ being coprime in this range. Then 
$x_n =f(x_{n-1})/x_{n-2}\equiv f(0)/x_{n-2}\equiv 1/x_{n-2}\bmod x_{n-1}$, 
so $x_{n+1}x_{n-1}=f(x_n)$ and then 
using (\ref{recip}) and (\ref{prods}) once more 
gives $f(x_n)\equiv f(1/x_{n-2})\equiv \pm 
x_{n-2}^{-d}\,f(x_{n-2})\equiv \pm x_{n-2}^{-d}\,x_{n-3}\,x_{n-1}\equiv 0 
\bmod x_{n-1}$, whence $x_{n-1}|f(x_n)$ and 
$x_{n+1}\in\ring$ as required. Furthermore, 
suppose that $p$ is an irreducible element of $\ring$ such that 
$p|x_{n+1}$; then $p|f(x_n)$ and $f(x_n)\equiv 1 \bmod x_n$, so $p\not |x_n$ 
and hence $x_n$ and $x_{n+1}$ are coprime, which completes the inductive 
step. This proves that $x_n\in\ring$ for all positive indices $n$, and 
the result extends to negative $n$ by the reversibility of (\ref{prods}). 

As for singularity confinement, note that a singularity can 
only occur in (\ref{prods}) if one of the iterates becomes zero. So 
let $x_{n-3}=a$ and $x_{n-2}=r+\ep$ where $r$ is any one of the (generically 
distinct) roots of $f$. Thus $x_{n-1}=a^{-1}\,f'(r)\,\ep +O(\ep^2)\to 0$ 
as $\ep\to 0$, while $x_n=f(0)/r +O(\ep )=1/r+O(\ep )$, which 
gives 
$x_{n+1}=(f'(r))^{-1}\, a \,\ep^{-1}f(1/r+O(\ep ))=O(1)$ because $f(r)=0$ 
implies $f(1/r)=0$ by (\ref{recip}), and thereafter all 
the terms are finite (and non-zero) as $\ep\to 0$. So we see that 
the singularity is confined  
at the first possible stage  
for any mapping of the form  (\ref{prods}) with $f(0)\neq 0$ that 
has the Laurent property. Conversely, it is easy to see that if we 
start 
from  
(\ref{prods}) with a polynomial $f$ such that $f(0)\neq 0$, 
we can always scale so that $f(0)=1$, and if we require that    
a zero is confined  immediately, then for any root $r$ of $f$, the 
reciprocal value $1/r$ 
must also be a root, which implies that (\ref{recip}) must hold,  
and hence the mapping has the Laurent property. 

In the above considerations we imposed the restriction 
$f(0)\neq 0$. We now describe the situation for the case $f(0)=0$, and 
so set $f(x)=x^M \, F(x)$ with integer $M\geq 1$, $F(0)\neq 0$ and 
$\deg F=D\geq 1$. 
There are two further cases to consider, according to whether 
$M=1$ or $M\geq 2$. In case (ii), we find that for the Laurent 
property to hold we must have 
\beq \label{mone} 
f(x)=x\, F(x), \qquad F(x)=\pm x^D\, F(1/x) 
\eeq 
for all $x$, with $F(0)=1$ or $-1$ for $D\neq 1$. 
So $F$ must satisfy the 
same reciprocal property as for $f$ in the generic case, and once again 
this condition is also sufficient; the proof of the Laurent property in 
this case is slightly more involved and will be presented 
elsewhere. Similarly to the argument for (\ref{recip}), 
it is easy to verify 
that the condition (\ref{mone}) 
also implies that the mapping (\ref{prods}) passes the singularity 
confinement test.  The other case (iii) 
is somewhat different, for upon taking 
\beq \label{mgtwo} 
f(x)=x^M\, F(x), \qquad M\geq 2, \qquad F \,\, \mathrm{arbitrary}, 
\eeq      
we find that the mapping defined by (\ref{prods}) gives 
$x_n\in\ring$ for all $n$. To see this, take as the inductive 
hypothesis that $x_{n-1}$, $x_n$ and $\rho_n=(x_n/x_{n-1})\in\ring$, and then 
write $\rho_{n+1}=(x_{n+1}/x_n)=\rho_n\, x_n^{M-2}\, f(x_n)\in\ring$ 
by the hypothesis, so $x_{n+1}=\rho_{n+1}\, x_n \in \ring$ as required. 
However, to see if a zero is confined in such a mapping we 
set $x_n =a$, $x_{n+1}=r+\ep$ where $F(r)=0$, so that 
$x_{n+2}=C_2\,\ep +O(\ep^2)\to 0$ 
as $\ep\to 0$, where $C_2=a^{-1}r^M\,F'(r)$. Subsequent terms have 
$x_{n+3}\sim C_3\,\ep^M$, $x_{n+4}\sim C_4\,\ep^{M^2-1}$, 
$x_{n+5}\sim C_5\, \ep^{M^3-2M}$ etc. for certain (non-zero) 
constants $C_j$, so the powers of $\ep$ continue to grow and   
the zero is \textit{not} confined. (However, note that we have explicitly 
excluded the 
trivial case $F=\be =\mathrm{constant}$, $f(x)=\be x^M$, which 
always has the Laurent property and satisfies singularity confinement in the 
sense  that $x=0$ cannot be reached from any non-zero initial data.)   

We have seen that there is a close connection between the Laurent 
property and singularity confinement for discrete equations 
of the form (\ref{prods}), 
but what about the \textit{integrability} of such mappings? 
For a measure-preserving mapping of the plane to be integrable it must 
have a conserved quantity. We can start by considering the lowest degree 
examples of polynomials $f$. If $d=1$ then 
\beq \label{done} 
x_{n+1}\,x_{n-1} =\ga \,x_n + \de, 
\eeq 
and any such map is integrable because it has a 
conserved quantity $K$ defining an elliptic curve, i.e. $K$ equals 
\beq \label{k} 
x_{n-1}+x_n+\frac{ \ga     
( x_{n-1}^2 +x_{n}^2 )   
+(\de +\ga^2)(
x_{n-1}+x_{n})
+\ga\de}{x_{n-1}x_n}.
\eeq 
Equation (\ref{done}) is 
bilinearized via $x_n=\tau_{n+3}\tau_{n-2}/(\tau_{n+1}\tau_n)$, yielding 
the special Somos-7 recurrence  
$$\tau_{n+4}\,\tau_{n-3}=\ga \,\tau_{n+3}\,\tau_{n-2} 
+\de \, \tau_{n+1}\,\tau_n,$$  
which can be related to 
a Somos-5 recurrence and thence 
solved in terms of elliptic sigma functions \cite{honetams}. 
However, if we require that (\ref{done}) itself should possess 
the Laurent property, then either $\de =0$ with $\gamma$ arbitrary and the 
map cycles with period 6, or we are in the situation (\ref{recip}) so 
that $\ga=\pm 1$ with $\de=1$ and the map cycles with period 5. It is 
interesting to note that in the latter case (fixing $\ga =1$) this map 
is equivalent to 
the functional relation that appears in the 
thermodynamic Bethe ansatz for an $A_2$ scattering theory \cite{ysystems}. 

The case $d=2$ is special: if $\la=f(0)\neq 0$ then 
the Laurent property requires (\ref{recip}) to hold, and there is 
the extra freedom to leave $\la\neq 0$ arbitrary. 
Then there are two choices of sign, giving either  
\beq 
\label{conmap} 
x_{n+1}\,x_{n-1} =x_n^2 + \nu \, x_n +\la ,
\eeq
or alternatively $x_{n+1}\,x_{n-1} =-x_n^2 +\la$. The second choice 
can be transformed  into the first by taking $x_n=\ka_n\,\tilde{x}_n$ where 
$\ka_n^2=1$ and $\ka_{n+1}\ka_{n-1}=-1$ for all $n$, so we can just consider
(\ref{conmap}) which has a conserved quantity  that defines a  conic, namely 
\beq \label{conic} 
L=\frac{x_{n-1}}{x_n}+\frac{x_{n}}{x_{n-1}} 
+\nu \left(\frac{1}{x_{n-1}}+\frac{1}{x_{n}}\right)
+\frac{\la}{x_{n-1}x_n}.  
\eeq 
Furthermore, the iterates also satisfy a 
linear recurrence of the form (\ref{addmap}), viz.  
$$x_{n+1}+x_{n-1}=Lx_n-\nu ,$$    
so in terms of the initial data we have $L=L(x_0,x_1)\in\ring$ 
and also $x_n\in \Z [\nu ,x_0,x_1, L]\subset 
\ring$ for all $n$, which is even stronger than the Laurent 
property. When $d=2$ and $f(0)=0$ there are further sub-cases. If 
(\ref{mone}) holds then (up to rescaling $x_n \to \nu^{-1}x_n$) 
either $f(x)=x(x+1)$, which is just a special 
case of  (\ref{conmap}), or we have the opposite choice of sign 
and the mapping is given by  
\beq
\label{ratmap}
x_{n+1}\,x_{n-1} =-x_n( x_n + 1)  ,
\eeq
which is also integrable, having  the conserved quantity 
$$ 
\tilde{J}= 
\frac{x_{n-1}^4+ x_n^4 + (x_{n-1}-x_n)^2(2x_{n-1}+2x_n+1)}  
{x_{n-1}^2 
x_n^2}.  
$$ 
Moreover, for (\ref{ratmap}) 
the iterates also satisfy the sixth order     
linear recurrence 
$$x_{n+6}+(\tilde{J}-1)(x_{n+4}-x_{n+2})-x_n=0,$$  
which provides an alternative proof that $x_n\in\ring$  based on the 
the fact that 
$\tilde{J} \in \ring$ and $x_j\in \ring $ for $j=0,1,\ldots,5$.     

For $d\geq 3$ all of the maps (\ref{prods}) with the Laurent 
property are non-integrable. To see this, one can 
count the growth of degrees to show that the algebraic entropy 
is non-zero \cite{hv}, but a simpler test to apply is Halburd's 
Diophantine integrability criterion \cite{halburd}. For a rational 
number $x=p/q$ the logarithmic height is $h(x)=\log \max \{ |p|,|q| \}$, and 
Halburd's test requires that for rational-valued maps to be integrable, 
$h(x_n)$ must grow no faster than 
a polynomial in $n$. 
Suppose that a sequence of 
(real or complex) iterates is such that $|x_n|\to\infty$ and 
$\Lambda_n=\log |x_n|\sim C\zeta^n$ for real $\zeta >1$ 
and some $C>0$. 
Then taking logarithms of both sides of (\ref{prods}) gives 
$ 
\Lambda_{n+1}+\Lambda_{n-1}- d\, \Lambda_n \approx 0   
$ and hence with $d\geq 3$ we find 
$\zeta =( d +\sqrt{d^2-4})/2>1$ as required. Now for  
equations (\ref{prods}) having the Laurent property, if we 
take $f$ to have integer coefficients and set $x_0=x_1=1$ then 
all $x_n$ are integers, so that $\Lambda_n=h(x_n)$. 
Moreover, if all the coefficients of $f$ are positive 
then the terms of the 
integer sequence will have precisely these asymptotics, so that 
the logarithmic height grows exponentially and 
$\lim_{n\to\infty}(\log h(x_n))/n=\log \zeta$ 
(which also happens to be the value of the algebraic 
entropy for these maps). 

To understand the deep connection between singularity 
confinement and the Laurent property, we propose to  
extend the above results in at least three 
directions.  
Firstly, given a pair of polynomials $f_1$, $f_2$ 
each satisfying (\ref{recip}),  
it is simple to prove that the composition $\phi_1\cdot \phi_2$  
of the corresponding pair of maps also has the Laurent property; 
the choice $f_1(x)=x^b+1$, $f_2(x)=x^c+1$  
generates a cluster algebra of rank 2 \cite{shz}. Secondly, 
there are many higher order discrete equations (integrable and 
non-integrable) with the Laurent property. Thirdly, this property 
can also apply to non-autonomous equations, 
such as the bilinear forms of discrete Painlev\'e equations.  

\noindent {\bf Acknowledgements.} Shortly after completing 
this work I learned from reading Gregg Musiker's Bachelor's thesis 
that  in 2001 David Speyer had also identified the case (i) with 
$f(0)\neq 0$, $f(x)=\pm x^d \, f(1/x)$ as being necessary 
and sufficient for the recurrence (\ref{prods}) 
to have the Laurent property; Speyer's unpublished proof (along 
very similar lines to the above) is reproduced in 
\cite{musiker}.  
I am grateful to Nalini Joshi for suggesting that I 
should consider maps of the form (\ref{prods}), and would 
like to thank  
Jim Propp for useful comments on an earlier draft. I also  
acknowledge the support of the Engineering 
$\&$ Physical Sciences Research Council.

\singlespacing

\end{document}